\documentclass[a4paper]{article}

\usepackage{titling}
\usepackage[centertags]{amsmath}
\usepackage{chicago,fancyhdr}
\usepackage{amssymb,amsfonts,eucal,oldgerm}
\usepackage[text={6in,8.25in},centering]{geometry}
\usepackage[polutonikogreek,french,german,english]{babel}

\usepackage[colorlinks=true,citecolor=red,linkcolor=blue,urlcolor=blue,pdftex]{hyperref}

\newcommand{\skipline}[1][1]{\vspace*{#1\baselineskip}}

\newtheorem{theorem}{Theorem}[subsection]

\newtheorem{defn}[theorem]{Definition}

\newtheorem{prop}[theorem]{Proposition}

\newcommand{\resetcounters}{%
  \setcounter{equation}{0}%
  \setcounter{theorem}{0}%
  \setcounter{figure}{0}%
}

\newcommand{\resetsec}{%
  \resetcounters%
  \renewcommand{\thetheorem}{\arabic{section}.\arabic{theorem}}%
  \renewcommand{\theequation}{\arabic{section}.\arabic{equation}}%
}
\newcommand{\resetsub}{%
  \resetcounters%
  \renewcommand{\thetheorem}{\arabic{section}.\arabic{subsection}%
    .\arabic{theorem}}%
  \renewcommand{\theequation}{\arabic{section}.\arabic{subsection}%
    .\arabic{equation}}%
  \renewcommand{\thefigure}{\arabic{section}.\arabic{subsection}%
    .\arabic{figure}}%
}

\renewcommand{\thetheorem}{\arabic{section}.\arabic{theorem}}%
\renewcommand{\theequation}{\arabic{section}.\arabic{equation}}%
\renewcommand{\thefigure}{\arabic{section}.\arabic{figure}}%

\newcommand{\french}[1]{\selectlanguage{french}#1\selectlanguage{english}}

\newcommand{\coloneq}{\mathrel{\mathop:}=}

\newcommand{\teog}{\text{\french{\emph{\og\hspace*{-.5ex}}}}}
\newcommand{\tefg}{\text{\french{\emph{\hspace*{-.5ex}\fg}}}}

\newcommand{\half}{\frac{1}{2}}

\newcommand{\athird}{\frac{1}{3}}

\title{Measure, Topology and Probabilistic Reasoning in
  Cosmology\thanks{I thank the members of the working group on
    Cosmology and Probability at the 2014 MCMP/Lausanne Summer School
    on Probability in Physics for invaluable feedback on a
    presentation of an earlier version of this paper, especially
    Jeremy Butterfield, Juliusz Doboszewski, Sam Fletcher, Karim
    Th\'ebeault and Jos Uffink.  I particularly thank Juliusz for
    catching errors in an earlier draft.  I thank Gordon Belot for
    pointing out ambiguities, infelicities, and unresolved technical
    issues, and for passing on to me some excellent stray thoughts.  I
    thank David Malament for pointing out a serious technical lacuna
    in the discussion of \S\ref{sec:inf-dim}.}}

\author{Erik Curiel\thanks{\textbf{Author's address}: Munich Center for
    Mathematical Philosophy, Ludwigstra{\ss}e 31,
    Ludwig-Maximilians-Universit\"at, 80539 M\"unchen, Germany;
    \textbf{email}: \href{mailto:erik@strangebeautiful.com}
    {\texttt{erik@strangebeautiful.com}}}}

\begin{document}
\maketitle

\skipline

\begin{quote}
  \begin{tabbing}
    \hspace*{8em}\=When every year  \=\kill
    \>A man said to the universe: \\
    \>``Sir, I exist!'' \\
    \>``However,'' replied the universe, \\
    \>``The fact has not created in me \\
    \>A sense of obligation.'' \\
    \>\>--- Stephen Crane
  \end{tabbing}
\end{quote}

\skipline

\begin{quote}
  \begin{center}
    \textbf{ABSTRACT}
  \end{center}  

  I explain the difficulty of making various concepts of and relating
  to probability precise, rigorous and physically significant when
  attempting to apply them in reasoning about objects
  (\emph{e}.\emph{g}., spacetimes) living in infinite-dimensional
  spaces, working through many examples from cosmology.  I focus on
  the relation of topological to measure-theoretic notions of and
  relating to probability, how they diverge in unpleasant ways in the
  infinite-dimensional case, and are difficult to work with on their
  own as well in that context.  Even in cases where an appropriate
  family of spacetimes is finite-dimensional, however, and so admits a
  measure of the relevant sort, it is always the case that the family
  is not a compact topological space, and so does not admit a
  physically significant, well behaved probability measure.  Problems
  of a different but still deeply troubling sort plague arguments
  about likelihood in that context, which I also discuss.  I conclude
  that most standard forms of argument used in cosmology to estimate
  the likelihood of the occurrence of various properties or behaviors
  of spacetimes have serious mathematical, physical and conceptual
  problems.
\end{quote}

\skipline

\thispagestyle{empty}
\tableofcontents

\section{Probabilistic Reasoning in Cosmology}
\label{sec:prob-reas-cosmo}

There is, by any standard measure, exactly one actual cosmos, and its
evolution cannot be repeated.  It is, therefore, perhaps surprising
when one first learns that probabilistic reasoning of various kinds
pervades cosmology as a science---reasoning not just about the
statistics of repeated and repeatable subsystems of the cosmos, but
reasoning that purports to assign probabilities to uniquely global
properties and structures of the cosmos itself.  It should, therefore,
perhaps not be surprising that problems arise for probabilistic
reasoning in this context peculiar to it.  

Physicists and philosophers have tended to focus on problems with
probabilistic reasoning in cosmology that, in the end, boil down to
one of the following two forms.
\begin{enumerate}
    \item What can probability mean, when there is only one physical
  system of the type at issue to observe?
    \item How can one justify attributions of definite values of
  probability when one cannot measure frequencies (because one cannot
  repeat experiments), which is to ask, what kinds of evidence may be
  available to try to substantiate attributions of probability?
\end{enumerate}
I shall not address these sorts of questions and problems in this
paper.\footnote{See \citeN{ellis-issues-phil-cosmo} and
  Smeenk~\citeyear{smeenk-logic-cosmo-revis,smeenk-phil-cosmo} for
  excellent reviews and discussion of these questions and problems.}
I shall rather address the relationship between topological and
measure-theoretic methods in probabilistic reasoning and the problems
that arise for it in the case of infinite-dimensional spaces, as
naturally occur in cosmology.

Although it is far more common to associate the mathematical theory of
measure spaces with probabilistic notions and reasoning, if one takes
a broad-minded view of what counts as ``probabilistic'' reasoning,
then, in many areas of physics, topological concepts and methods
ground much of what it is reasonable to think of as probabilistic
reasoning.  This is particularly true in a science such as cosmology,
in which well defined probability measures over families of systems
are few and far between.  In such situations, physicists often argue
that a property or behavior of interest is typical or generic or
stable in a family of possible systems, or is scarce or meagre or
rigid, and so on, with no serious attempt to make those ideas
quantitatively precise, though they clearly are intended to have
probabilistic import.  Often, the arguments are grounded on
topological considerations with gestures at interpreting the
conclusions in measure-theoretic terms so as to justify the intended
probabilistic import.

Say we are interested in the likelihood of the appearance of a
particular feature (having a singularity, \emph{e}.\emph{g}.\@) in a
given family of spacetimes satisfying some fixed condition (say, being
spatially open).  If one can convincingly argue that spacetimes with
that feature form a ``large'' open set in some appropriate, physically
motivated topology on the family, then one concludes that such
spacetimes are generic in the family, and so have high prior
probability of occurring.  If one can similarly show that such
spacetimes form a meagre or nowhere-dense set in the family, one
concludes they have essentially zero probability.  The intuition
underlying the conclusions always seems to be that, though we may not
be able to define it in the current state of knowledge, there should
be a physically significant measure consonant with the topology in the
sense that it will assign large measure to ``large'' open sets and
essentially zero measure to meagre or nowhere-dense sets.  Similarly
for stability and rigidity: if one can show that a given feature is
topologically stable under ``small'' perturbations, one can conclude
that the probability is very high that a spacetime approximately
satisfying the relevant conditions will still have the feature; if the
feature is topologically rigid under ``small'' perturbations, one can
conclude that the probability is essentially zero that a spacetime
approximately satisfying the relevant conditions will still have the
feature.  In order to justify the probabilistic nature of the
conclusion, one again assumes the existence of an appropriate measure
consonant with the topology in the sense that the smallness of the
pertubation is to be judged by the fact that the resulting spacetime
is in a neighborhood of the initial spacetime, of ``small'' measure.

In cosmology, reasoning of this form occurs ubiquitously, in the
context of the following kinds of problem:
\begin{enumerate}
    \item characterizing the likelihood of observing certain kinds of
  events, given the situation of possible observers in a spacetime,
  \emph{i}.\emph{e}., the fact that observers are limited in
  observations to what lies in their past light-cone, by the
  sensitivity of their apparatus, by the amount of time a process will
  emit energy of a given magnitude or greater, and by how far to the
  past of the observers such processes may occur
    \item characterizing the likelihood that the value of a universal
  constant lies within a fixed range
    \item characterizing the likelihood that cosmological initial
  conditions of a particular kind or form, or having a particular
  property or characteristic, obtained
    \item characterizing the likelihood that large-scale structure of
  a particular kind would form
    \item characterizing the likelihood that a spacetime has a
  particular global (causal, topological, projective, conformal,
  affine, metrical) property
\end{enumerate}
Common specific examples of such problems are:
\begin{enumerate}
    \item characterizing the likelihood that observers such as
  ourselves would come to exist in the sort of spatiotemporal region
  we occupy in a spacetime of this sort
    \item characterizing the likelihood that we are ``typical''
  observers in the universe
    \item characterizing the likelihood that the cosmological constant
  has any non-zero value, and has, moreover, a value near that
  actually observed
    \item characterizing the likelihood of various ``fine-tuning
  coincidences'': the seeming equality of densities of dark energy and
  dark matter in the current epoch; the approximate flatness of the
  observed universe; the approximate isotropy and spatial homogeneity
  of the observed universe; the seemingly required special entropic
  state of the very early universe; \emph{etc}.
    \item characterizing the likelihood that a spatially open
  spacetime is future-singular
\end{enumerate}

In most branches of physics, one would address such problems by fixing
an appropriate reference class of physical systems and a physically
significant probability-measure on that class.  When one cannot
rigorously define such a measure, or one is not that interested in
quantitative exactness, one will often rest content with arguing (or
just stipulating) that a physically significant measure exists whose
distribution of weight harmonizes in a particular way with a natural
topology on the class of systems, to wit, one assumes that non-trivial
positivity of measure is at least strongly correlated with openness of
sets and likewise that smallness or nullness of measure is correlated
with topological meagreness of sets.  In this case, one will base
one's estimates of likelihood on the topological properties of the
families of systems at issue.

Even in cases where one does have a well defined measure to give
quantitative exactness to estimates of genericity or typicality,
however, one still needs the measure to harmonize with an underlying
reasonable topology in the appropriate way.  The point is simple,
though it does not seem to be widely appreciated or even recognized,
either in the physics or the philosophy literature: genericity and
typicality, roughly speaking, mean something like ``most systems are
similar in this respect'' (and \emph{mutatis mutandis} for meagreness
and scarcity); ``most'', however, is a measure-theoretic notion,
whereas ``similar in this respect'' is a topological
notion.\footnote{One can of course quantify similarity using a metric
  as well, but in this case the metric will give rise to a topology.
  The measure will still have to harmonize with the metric and so will
  automatically harmonize (or not) with the induced topology.}  Most
systems satisfy a property if the family of such systems forms a set
of large measure; a given family of physical systems are similar in a
given respect if the topological neighborhood-systems of the elements
of the space representing the physical systems stand in some
appropriate relation to each other, which often will be as simple as
the fact that the family of elements representing the physical systems
forms an open set.

In cosmology, however, the systems one most often focuses on are
entire spacetimes, and families of spacetimes usually form
infinite-dimensional spaces of a particular kind.  And now one comes
to the heart of the problem: it is a theorem (as I discuss in some
detail in \S\ref{sec:inf-dim} below) that infinite-dimensional spaces
of that kind do not admit non-trivial measures that harmonize in the
right way with any underlying reasonable topology.  It follows that
one simply does not have available the kinds of reasoning normally
employed to draw even qualitative conclusions about the likelihoods of
properties or features or behaviors of spacetimes.  To be clear, I do
not claim that it is not possible to draw well grounded conclusions
about such likelihoods, only that arguments of the standard forms
cannot, not even in principle, be made rigorous, and so conclusions
based on them are \emph{prima facie} suspect, and should be treated
with far more caution and skepticism than is common in the physics and
philosophy literature.  It is exactly the standard forms of argument,
however, that cosmologists make when reasoning about likelihoods.

In \S\ref{sec:topo-meas-prob}, I quickly review the basics of
topology, measure theory and probability theory, emphasizing technical
and interpretative points that the rest of the paper relies on.
Cognoscenti may want to skip that section, though I do discuss some
issues (such as the character of topologies on spaces of functions,
and the topological character of the uniqueness of the Lebesgue
measure on $\mathbb{R}^n$) sometimes unfamiliar even to those with a
solid grounding in topology and measure theory.  I also present the
basic facts about topology in a somewhat unusual way, based on the
idea of an accumulation point, which is particularly suited to the
goals of this paper.  In my presentation of the basics of probability,
moreover, I focus on those foundational problems most relevant to the
kinds of cosmological argument I examine.  In \S\ref{sec:inf-dim}, I
briefly rehearse the relevant aspects of topology and measure theory
in the context of infinite-dimensional Fr\'echet spaces, and conclude
with a statement of the fundamental theorem relevant to this paper and
explain its import.  In \S\ref{sec:topo-sts}, I discuss the few well
defined topologies on families of spacetimes commonly used in
cosmology, and show that they have severe problems of physical
interpretation on their own.  In \S\ref{sec:meas-sts}, I do the same
for the only known example of a well defined measure on a
finite-dimensional family of spacetimes of real physical interest.  I
conclude in \S\ref{sec:gener-scarc-pred} with a discussion of several
standard cosmological arguments about likelihood in the context of
infinite-dimensional spaces of spacetimes, and show how the reasoning
runs afoul of the mismatch between topology and measure in such
spaces.

\section{Topology, Measure, Probability}
\label{sec:topo-meas-prob}

\subsection{Topological Spaces}
\label{sec:topo}
\resetsub

A topology $\mathfrak{T}$ is a family of sets, including the null set
$\varnothing$, closed under arbitrary unions and finite
intersections.\footnote{All the material I cover in this section is
  developed with thoroughness and illuminating insight in
  \citeN{kelley-gen-topo}.}  In particular, the union $\mathcal{T}$ of
all elements of $\mathfrak{T}$ itself belongs to $\mathfrak{T}$, and
is called the topological space with topology $\mathfrak{T}$.  The
elements of $\mathfrak{T}$ are its open sets; a neighborhood of a
point of $\mathcal{T}$ is a subset of $\mathcal{T}$, not necessarily
in $\mathfrak{T}$, that contains an open set containing that point.
In general, one can associate many different topologies with the same
set of points $\mathcal{T}$.  (We will, however, still abuse notation
and terminology in the usual way when no ambiguity can arise,
sometimes referring to a topological space simply by its associated
set without specifying which topology on it we mean.)

As is always the case with mathematical fields of study, there are
many ways to think about the subject of topology, both in the sense of
intuitive visualization and in the sense of rigorous formalization.
For our purposes, the sense in which topology captures the idea of the
study of ``continuity''---what remains invariant under deformations of
a space that don't rip or puncture it and don't glue different parts
together---is the most important.\footnote{A topologist is a person
  who doesn't know the difference between a coffee-cup and a
  doughnut.}  The neighborhoods of a topology capture an idea of
relative proximity relevant to the idea of continuity: two points of
the underlying set are in proximity (relative to the fixed topology)
if the family of neighborhoods of one stands in one of a number of
relations to the family of neighborhoods of the other.  Intutively
speaking, a neighborhood is a region of the space in which, at the
point of which it is a neighborhood, ``arbitrarily small
perturbations'' don't take one out of the region.  If one thinks of
the topology as capturing something like a similarity relation among
entities, then a neighborhood of an entity is a collection of other
entities similar to the first to some degree.

For our purposes, one of the most important of these relations among
families of neighborhoods is grounded on the idea of an accumulation
point.  Given any subset $O \subset \mathcal{T}$ (whether an open set
or not), an \emph{accumulation point} of $O$ is a point $p$ such that
every neighborhood of $p$ has a non-trivial intersection with $O -
\{p\}$.  In agreeably suggestive language, one may say that an
accumulation point is ``arbitrarily close'' to its associated set.
Much information about the topology of a topological space is encoded
in the behavior of infinite sequences of points, and in particular by
the situation of any accumulation points they may have.  (Indeed,
under mild restrictions, which all the examples we consider here
satisfy, a topology can be fully characterized by the behavior of the
accumulation points of all infinite sequences.)  A set is
\emph{closed} if it contains all its accumulation points.

Given a sequence $P = \{p_i\}$ ($i \in \mathbb{I}^\uparrow$, the
non-negative integers), we say $P$ is \emph{eventually in} a set $O$
if there is an $m \in \mathbb{I}^\uparrow$ such that $p_n \in O$ for
all $n > m$.  Clearly, if there is a $p$ such that a sequence $P$ is
eventually in every one of its neighborhoods, then $p$ is an
accumulation point of $P$.  In this case, we say $P$ \emph{converges}
to $p$.\footnote{A sequence may have an accumulation point it does not
  converge to.  A sequence is \emph{frequently in} a set $O$ if, for
  every $m \in \mathbb{I}^\uparrow$, there is an $n > m$ such that
  $p_n \in O$.  If a sequence is frequently in every neighborhood of a
  point, that point is a \emph{cluster point} of a sequence.  A
  cluster point is an accumulation point; a sequence may, but does not
  necessarily, converge to a cluster point.  Roughly speaking, a
  sequence may ceaselessly approach arbitrarily close to and then
  recede from a cluster point, but never come to remain permanently
  near it.}  If a sequence converges at all, there may be more than
one point the sequence converges to, depending on global properties of
the topology.  A topology is \emph{Hausdorff} if every two distinct
points have disjoint neighborhoods.  In a Hausdorff space, if a
sequence converges, its convergence point is unique.

A function from one topological space to another is \emph{continuous}
if the inverse image of an open set in the range is an open set in the
domain: if you tell me how proximate you want to be to a point in the
range, under the mapping, I'll tell you how proximate you need to be
to its pre-image in the domain.  Under mild conditions on the
topology, the continuity of a function can be characterized by the
behavior of infinite sequences of points: roughly speaking, a function
$f$ is continuous if, for every sequence $P$ in the domain that
accumulates at a point $p$, the sequence $f[P]$ in the range
accumulates at $f(p)$.  

Whether or not a given mapping between two point-sets is continuous
depends sensitively on the topologies one imposes on the sets.  As a
general rule, the fewer open sets a topology has, the easier it is for
a function having the space as its range to be continuous; contrarily,
the fewer open sets a topology has, the harder it is for a function
having it as its domain to be continuous.  The intuition behind this
rough claim is easy to grasp: the more open sets there are, the harder
it is for a sequence to have an accumulation point.  Of two topologies
on a given set, one is \emph{finer} than the other if every one of its
open sets is also an open set of the other.  (One also says that the
other is \emph{coarser} than the one.)  Finer topologies have more
continuous functions from them; coarser topologies have more
continuous functions to them.

One of the most central and important ideas in topology is
compactness.  The motivation behind the idea comes from the classic
Heine-Borel Theorem.  To state it, we need two more definitions.  An
\emph{open cover} of a subset of a topological space is a family of
open sets whose union contains the subset.  A \emph{subcover} of a
cover is a subset of the cover that is also itself a cover.
\begin{theorem}[Heine-Borel]
  \label{thm:heine-borel}
  Every open cover of a closed, bounded interval of $\mathbb{R}$
  (under its standard topology) has a finite subcover.
\end{theorem}
This is remarkable.  No matter how large and fiendishly Baroque one
makes an open cover of a closed, bounded interval, one can
\emph{always} select a finite number of elements from it that will
still cover the interval.  As with all the best theorems, the
conclusion of the Heine-Borel Theorem has become a definition of
fundamental importance: a subset of a topological space is
\emph{compact} if every one of its open covers has a finite subcover.
(Of course, the entire space itself may be compact.)

Compact sets have particularly pleasant properties for our purposes,
perhaps the two most important of which are that, first, under mild
restrictions on the topology, every infinite sequence in a compact set
has at least one accumulation point, and, second, under no
restrictions at all on the topology, the Cartesian product of any
family of compact spaces is itself compact under the natural product
topology.  (The latter statement is known as Tychonov's Theorem.)
Intuitively speaking, then, compact sets don't ``extend out to
infinity'', and they also contain ``every point they could possibly
have had in the first place''---in a natural sense, they are bounded,
and they don't have any gaps or holes.  An important weakening of the
notion of compactness retains almost all its nice properties: a
topological space is \emph{locally compact} if every point has a
compact neighborhood.

Finally, we record a few definitions and propositions that will play
an important role in what follows.  A subset $D$ of a topological
space is \emph{dense} if every point of the space is an accumulation
point of some sequence of points in $D$.  Intuitively, $D$ extends
arbitrarily closely to every point of the space.  The rational
numbers, for example, form a dense subset of $\mathbb{R}$ (indeed, a
countable one).  A topological space is \emph{separable} if it has a
countable dense subset.  A subset of a topological space is
\emph{nowhere dense} if the union of the set and all its accumulation
points do not contain an open set.  If a subset $N$ is nowhere dense,
then, given any point not in $N$, one can find a neighborhood around
that point such that no sequence in $N$ accumulates on the
neighborhood.

The case of most interest for us will be topologies on the family of
continuous, differentiable or smooth functions between two topological
spaces---in particular, the family of cross-sections on the fiber
bundle of Lorentz metrics over a candidate spacetime manifold
(connected, paracompact, Hausdorff, four-dimensional).\footnote{A
  topological space is \emph{connected} if it is not the union of two
  open, nonempty, disjoint sets.  The exact definition of
  paracompactness is too involved to give here; suffice it to say that
  it means the space is not ``too big''.  Indeed, one has to work hard
  to construct a topological space that is not paracompact
  \cite{hocking-young61}.  In any event, a theorem due to
  \citeN{geroch-spin-struc-gr-i} shows that a manifold has a Lorentz
  metric only if it is paracompact, so we lose nothing by restricting
  attention to such manifolds.}  Consider two topological spaces
$\mathcal{T}_1$ and $\mathcal{T}_2$, and the family of continuous
functions $\mathfrak{F}$ from the former to the latter.  Define
$\mathcal{N} (K, \, O) \coloneq \{ f \in \mathfrak{F}: f[K] \subset
O$, for $K \subset \mathcal{T}_1$ compact and $O \subset
\mathcal{T}_2$ open$\}$.  The family of all such collections, for all
such $K$ and $O$, forms a subbase for the \emph{compact-open topology}
on $\mathfrak{F}$.\footnote{A \emph{base} for a topology is a
  collection of open sets such that every other open set can be formed
  from a union of sets in the base.  A subbase is a collection of open
  sets such that one can form a base by taking finite intersections of
  them.}  A topology on $\mathfrak{F}$ is said to be \emph{jointly
  continuous} if the mapping $P : \mathfrak{F} \times \mathcal{T}_1
\rightarrow \mathcal{T}_2$ that takes $(f, \, p)$ to $f(p)$ is itself
continuous, in the product topology on $\mathfrak{F} \times
\mathcal{T}_1$.\footnote{The product topology for the Cartesian
  product of two topological spaces is exactly what one would expect:
  all sets of the form $O_1 \times O_2$, where $O_1$ is an open set in
  the first factor and $O_2$ open in the second, constitute a base for
  the product topology.}  Say a topological space $\mathcal{T}$ is
\emph{regular} if for every $p \in \mathcal{T}$ and every neighborhood
$N$ of $p$, there is a closed neighborhood $U$ of $p$ such that $U
\subset N$.  Then the following proposition captures the sense in
which the compact-open topology is the coarsest mathematically
reasonable topology to impose on a function space, so long as one
wants that topology to respect the structure of the elements of the
space \emph{as} functions.
\begin{prop}
  \label{prop:c-o-topo}
  If the topological space $\mathcal{T}$ is locally compact and
  regular, then the compact-open topology is the coarsest jointly
  continuous topology one can impose on the family of continuous
  functions from $\mathcal{T}$ to any other topological space.
\end{prop}

Most relatively well behaved topologies on spaces over which one
considers spaces of functions---and in particular all the ones we will
consider here---are Hausdorff, separable, regular, and locally
compact.

\subsection{Measure Spaces}
\label{sec:meas}
\resetsub

A \emph{$\sigma$-algebra} is an ordered pair $(\mathcal{S}, \,
\Sigma)$ consisting of a set $\mathcal{S}$, and a non-empty collection
of subsets of $\mathcal{S}$, $\Sigma$, closed under the operations of
finite set-differences and countable unions.\footnote{All the material
  I cover in this section is developed with thoroughness and
  illuminating insight in \citeN{halmos-meas-thry}.}  Where no
confusion can arise, we will often abuse notation in the standard way
and refer to $\Sigma$ itself as the $\sigma$-algebra.  Write
`$\mathbb{R}^\uparrow$' for the set of non-negative real numbers.
\begin{defn}
  \label{def:meas}
  A \emph{measure} on a $\sigma$-algebra $\Sigma$ is a function $\mu :
  \Sigma \rightarrow \mathbb{R}^\uparrow \cup \{\infty\}$ such that
  \begin{enumerate}
      \item $\mu (S) < \infty$ for at least one $S \in \Sigma$
      \item for any countable, pairwise-disjoint family $\{S_i\}
    \subset \Sigma$,
    \[
    \mu \left( \bigcup_i S_i \right) = \sum_i \mu (S_i)
    \]
  \end{enumerate}
\end{defn}
A \emph{measure space} is an ordered pair consisting of a
$\sigma$-algebra and a measure on it; the elements of the
$\sigma$-algebra are called \emph{measurable sets}.  It follows from
the definitions that the null set $\varnothing$ is always measurable,
and any measure assigns value zero to it.\footnote{While it is not
  unusual, it is also not entirely standard to demand condition 1 for
  a measure.  I do it because it simplifies matter greatly, in
  particular guaranteeing that the null set is measurable, of measure
  zero, without having to require it as a separate axiom.  Also, it
  seems to me a quite reasonable bare minimum one should require of
  something one wants to call a measure, if it is to be useful in
  physics at all.}  In general, however, the null set will not be the
only set assigned a measure of zero.  We say a property that holds for
all points of a measure space except for a subset of measure zero
holds \emph{almost everywhere}.

A $\sigma$-algebra has much the same feel about it as a topology,
naturally giving rise to the question whether one can construct
measures on topological spaces that relate in a natural way to the
topology.
\begin{defn}
  \label{def:borel-sets-meas}
  Let $\mathcal{T}$ be a Hausdorff compact topological space.  The
  \emph{Borel sets} $\mathcal{B}$ of $\mathcal{T}$ consist of the
  smallest $\sigma$-algebra containing all its open sets.  A
  \emph{Borel measure} is a measure $\mu$ on the Borel sets such that
  $\mu (C) < \infty$ for every compact set $C$.
\end{defn}
A Borel measure, in an obvious and natural sense, respects the
topology of the underlying topological space.\footnote{Again, while
  not unusual, it is not wholly standard to demand that a Borel
  measure assigns finite measure to all compact sets.  And, again,
  this seems to me the minimum one should require of such a thing for
  it to be usefully applicable in physics.}  It is a simple matter to
construct a natural Borel measure on $\mathbb{R}$: one is uniquely
picked out by the requirement that $\mu ([a, \, b]) = b - a$ for every
real interval $[a, \, b]$.  This measure suitably generalized to
$\mathbb{R}^n$ is not however unique, and its multiplicity can be
traced to the fact that it lacks one feature, completeness, that it is
convenient to have: we say a measure is \emph{complete} if every
subset of every set of measure zero is itself measurable (and so
necessarily of measure zero).

An \emph{extension} of a measure space $(\mathcal{S}, \, \Sigma, \,
\mu)$ is another measure space $(\mathcal{S}, \, \Sigma', \, \mu')$
such that $\Sigma \subset \Sigma'$ and $\mu'(A) = \mu(A)$ for all $A
\in \Sigma$.  The following is easily proven.
\begin{prop}
  \label{prop:uniq-exten-borel-meas}
  There is exactly one complete extension of all natural Borel
  measures on $\mathbb{R}^n$ (for any $n \in \mathbb{I}^\uparrow$).
\end{prop}
Lebesgue measure $\mu_\textsc{l}$ on $\mathbb{R}^n$ is the unique
complete extension of any of the natural Borel measures.  Any
countable subset of $\mathbb{R}^n$ has Lebesgue measure zero, but
uncountable subsets also can.  The Cantor Set is an example.

Let $A$ be a subset of $\mathbb{R}^n$; then, for any $p \in
\mathbb{R}^n$, denote by `$A + p$' the set that results by adding $p$
to every element of $A$, sometimes called the $p$-translate of $A$.
Now, the following proposition shows the most important properties of
Lebesgue measure in relation to the natural topology and linear
structure on $\mathbb{R}^n$.
\begin{prop}
  \label{prop:leb-meas-props}
  Lebesgue measure is locally finite, strictly positive and
  translation invariant, \emph{i}.\emph{e}.:
  \begin{enumerate}
      \item every $p \in \mathbb{R}^n$ has an open neighborhood $O$
    such that $\mu_\textsc{l} (O) < \infty$
      \item $\mu_\textsc{l} (O) > 0$ for every non-empty open set $O$
      \item for every measurable set $A$ and every $p \in
    \mathbb{R}^n$, $\mu_\textsc{l} (A + p) = \mu_\textsc{l} (A)$
  \end{enumerate}
\end{prop}
(Lebesgue measure is obviously not the unique measure satisfying these
conditions, because there are non-complete Borel measures that also
satisfy them.)  The translation invariance of Lebesgue measure is
commonly taken to be its most characteristic feature, to the point
that any translation-invariant measure on any linear space is often
referred to as a Lebesgue measure.  The analogous property is
particularly important in a measure that would be used to define a
probability space over a family of events that itself has an
appropriate algebraic structure, for reasons I discuss in
\S\ref{sec:prob} below.

The following theorem captures the precise sense in which Lebesgue
measure is the unique measure that respects both the topology and the
linear structure of $\mathbb{R}^n$.
\begin{theorem}
  \label{thm:uniq-leb-meas}
  Lebesgue measure is the unique complete, translation-invariant
  measure on the Borel sets in $\mathbb{R}^n$.
\end{theorem}
From hereon, we will consider only Borel measures that are, like
Lebesgue measure, strictly non-negative (\emph{i}.\emph{e}., ones that
assign negative values to no measurable set).  


\subsection{Probability}
\label{sec:prob}
\resetsec

Measures allow for one elegant and important way to formalize the
notion of probability: a \emph{probability space} is an ordered pair
consisting of a $\sigma$-algebra $(\mathcal{P}, \, \Pi)$ and a
strictly positive measure $\mu$ on it such that $\mu(\mathcal{P}) =
1$.  Intuitively speaking, the elements of $\mathcal{P}$ represent the
totality of possible outcomes for some family of phenomena we are
interested in, those of $\Pi$ the collections of outcomes to which it
makes sense to assign probabilities, and the value assigned by $\mu$
to an element of $\Pi$ the probability of that collection of outcomes.
It is trivial to show that a probability space satisfies the standard
Kolmogorov axioms of probability theory.

It is of fundamental importance to recognize that, when one wants to
be precise, clear and unambiguous, it never makes sense to ask for
``the probability'' \emph{simpliciter} of some event or collection of
events.  One must have in hand a probability space that includes the
event or collection of events in its $\sigma$-algebra (or, at least,
some structure formally equivalent to one).  In general, for any given
event or collection of events, there will be many, many, many such
probability spaces, with different $\sigma$-algebras and with
different measures.  Picking the most appropriate $\sigma$-algebra for
the question or investigation at hand is known as the
\emph{reference-class problem}.  I am not aware of any standard name
for the problem of picking the most appropriate measure, but it is
equally as important and difficult in general as the reference-class
problem.  The kinds of consideration that should bear on those choices
will depend on the nature of the subject matter one is treating, and
on the nature of the problem concerning that subject matter.  In
physics, of course, when using measures to assign probabilities to
collections of events, one wants to find a $\sigma$-algebra that
represents ``all appropriately similar events'', where the similarity
has manifest physical significance for the problem at issue, and to
fix a measure on it that captures a property of real physical
significance shared by the events that relates in a clear, direct,
determinate way to the probabilities one wants to
characterize.\footnote{Peirce~\citeyear{peirce-doct-chnc,peirce-prob-ind}
  gives a particularly beautiful discussion of these issues, although
  of course he does not use the language of measures.}  Without having
made sure that the measure latches on to and respects a physically
significant feature of the problem space with manifest relevance to
the determination of probabilities, there will be no reason to think
of the values the measure assigns as representing real physical
probabilities.  All these issues play a crucial role in attempts to
evaluate the soundness of many kinds of argument in cosmology.

Although measures on their own can be used to define probability
spaces, it is often the case that topological considerations play an
important role in probabilistic reasoning.  It is almost always
desirable, for instance, especially in physics, for an appropriate
probability measure to be a Borel measure, and in particular to assign
non-zero probability to any collection of outcomes that forms an open
set in a physically natural topology.  This captures the idea that, if
an event has non-zero probability (measure greater than zero), then
``arbitrarily small'' perturbations of it shouldn't render the result
impossible, \emph{i}.\emph{e}., send it into a set of measure zero.
One can guarantee this by having the original event lie in an open
set, which, because the measure is Borel, will have positive measure.
If this were not the case, then, given the necessarily limited
precision of observations in physics, we would find ourselves in the
position of predicting outcomes with non-zero probability that we
could never in principle observe.

Topology plays other important roles in probabilistic reasoning as
well.  In many cases, the quantitative exactness delivered by a
measure is either not feasible or not desirable.  Sometimes it is
enough merely to know that an event is very likely or not likely at
all, without attaching a quantitatively exact probability to it.
Let's say that we make a prediction for the dynamical evolution of a
system starting from a set of exact initial conditions.  We want to
know how likely it is that the system, if prepared with approximately
those initial conditions will evolve in approximately the predicted
way.  (Roughly speaking, this is known as the Hadamard stability
problem for the initial-value formulation of the system.)  One natural
way to make the question precise is to find appropriate topologies for
the space of initial conditions and the space of dynamical evolutions,
define the mapping taking initial conditions to dynamical evolutions,
and determine whether it is jointly continuous.  (``Do arbitrarily
small perturbations of the initial conditions leave the later
dynamical behavior essentially unchanged?'')  If so, then, if the
prediction is sound and if we have good reason to believe that the
system starts with initial conditions close enough to the exact
initial conditions used to generate the prediction, it is very likely
that we will get the expected behavior even though we know that, due
to the finite exactness of measurement and preparation, the system
almost certainly did not start with those exact initial conditions.
If the mapping is not jointly continuous, then it may be very unlikely
that we will get the expected behavior, no matter how close to the
exact initial conditions the system starts evolving from.  (This is
one of the reasons why it is almost always desirable, from a physical
point of view, to have one's function-space topology be jointly
continuous.)

In a closely related vein, say that we have found an appropriate
topology for the space representing the possible states of a type of
system, and that, moreover, the points of that space representing the
system as possessing a certain property with values in a fixed range
form an open dense subset.\footnote{If the topology arises from or is
  compatible with a complete metric, one can use the more general
  criterion that the set be a $G_\delta$-set, \emph{i}.\emph{e}., that
  it be a countable intersection of open dense subsets, in so far as
  the complement of a $G_\delta$-set in this case is nowhere dense.}
Then there is a natural sense in which it is overwhelmingly likely
that an appropriately random sampling of such systems will all evince
values for the given property falling within the fixed range.  If the
subset is nowhere dense, it is very unlikely to find a system having
the property with value in the fixed range in an appropriately random
sample.\footnote{If the topology arises from or is compatible with a
  complete metric, one can use the more general criterion that the set
  be \emph{meagre}, \emph{i}.\emph{e}., that it be a countable union
  of nowhere dense subsets, in so far as the complement of a meagre
  set in this case is dense.  (This is known as the Baire Category
  Theorem.)}  Of course, one must bear in mind that such conclusions
depend not only on the physical propriety of the topology one has
fixed on the space of states, but, at least as importantly, they
depend on the propriety of the mechanism one has chosen to construct
the random sample.  If one's sampling mechanism is biased in some way,
then it doesn't matter how ``appropriate'' one's measure is---one will
not get physically reliable results.  Again, these issues play a
crucial role in the evaluation of many kinds of cosmological argument.

Finally, one can use measures in similar ways to draw qualitative
judgments about the likelihood of an event or kind of event: a
property that holds almost everywhere in an appropriate measure space
will be very likely to occur, and one that holds in a set of measure
zero will be very unlikely, even if the measure is not a probability
measure, so long as one has been able to demonstrate an appropriate
relation between the measure and the relevant physical properties of
the system at issue.  It is only in the latter case that the value the
measure attributes to a set may reasonably be thought of as a
representation of a real physical probability for the class of events
in the set.

\section{Topology and Measure in Infinite-Dimensional Spaces}
\label{sec:inf-dim}
\resetsec

Now, as we have seen, there is a natural sense in which, in
$\mathbb{R}^n$, Lebesgue measure respects both the topology and the
linear structure, both of which are desirable features in a measure
one wants to found probabilistic reasoning on, for the reasons
discussed in \S\ref{sec:prob}, \emph{inter alia}.  The situation in
infinite-dimensional spaces, therefore, as we will see, poses
considerable problems for the hopeful physicist.  This matters
because, for our purposes---the application of probabilistic reasoning
to families of spacetimes---the natural spaces one works with are
spaces of functions (Lorentzian metrics on differential manifolds).
These spaces tend to be infinite-dimensional spaces with natural
(locally) algebraic structures accruing to them.

To apply probabilistic reasoning to families of spacetimes, one must
first choose what sort of spacetime metric one is going to work with,
$C^n$ or $C^\infty$.  Each has virtues and demerits.  To see what is
at issue in a simpler setting, consider the set of functions on the
unit disk.  For C$^n$ functions, we have the norm
\[
\|f\| = \sup |f| + \ldots + \sup |\nabla^{(n)} f|
\]
resulting in a Banach space, since this norm is complete for any
finite $n$---because the disk is compact and the functions are
continuous, the supremum is always finite.  (A \emph{Banach space} is
a normed vector space, complete with respect to the metric the norm
induces.)  For C$^\infty$, we do not get a Banach space because the
resulting infinite sum may not converge.  Instead, we define
\[
\teog f \tefg = \frac{\sup |f|}{1 + \sup |f|} + \ldots + \frac{1}{2^n}
\frac{\sup |\nabla^{(n)} f|}{1 + {\sup |\nabla^{(n)}} f|} + \ldots
\]
which manifestly converges.  This operation defines a metric in the
obvious way,
\[
(f,\, g) = \teog f - g \tefg
\] 
and this metric is complete in the sense that all its Cauchy sequences
converge to a point in the space.  This operation, however, does not
define a norm, since it is not the case that
\[
\teog af \tefg = |a| \teog f \tefg
\]
for scalar $a$.  (There is no norm for C$^\infty$ functions that both
accounts for all their derivatives and, in a natural sense, extends
the norm for C$^n$ functions.)  The metric is, however, manifestly
invariant with respect to translations.  The resulting space is a
\emph{Fr\'echet space}: a metrizable, locally convex vector space,
complete with respect to a translation-invariant metric.

Is it true in such a large space that every well-behaved vector field
has unique integral curves (a necessity, \emph{e}.\emph{g}., for
certain forms of local stability analysis)?  In a Banach space, yes,
but in a Fr\'echet space, no.  (Sometimes there are no integral
curves, sometimes they are not unique.)  As an example of the way
things can go awry, consider a map $\phi$ from functions on the disk
to functions on the disk defined as follows:\footnote{I thank Bob
  Geroch for conversations in which we worked out the details of this
  example.}
\[
\phi(f) = \xi^n \nabla_n f
\]
for $\xi^a$ a vector field on the disk.  $\phi$ wants to be a linear
functional on our space of functions, \emph{i}.\emph{e}., to be a
vector field on the vector space of functions on the disk, but, while
it is in fact a vector field on the Fr\'echet space of C$^\infty$
functions, it is not one on the Banach space of C$^n$ functions, since
for $f \in C^n$ it is not necessarily the case that $\nabla_a f \in
C^n$.  On the Fr\'echet space, however, the vector field resulting
from this mapping does not yield unique solutions---when one slides a
function along the disk, as the mapping in effect asks one to do, one
gets to make up whatever one wants to fill up the ``back'' part of the
disk, as the ``front'' part of the function slides off the disk,
\emph{i}.\emph{e}.,
\[
\frac{d}{dt}f(t,\, 0) = \xi^n \nabla_n f(t,\, 0)
\]
has no unique integral curves.  So one can have unique integral
curves, but no vector field (on Banach space), or a vector field but
no unique integral curves (on Fr\'echet space).

In general, however, there may be difficulties with defining tangent
vectors to curves on a Fr\'echet space.  In a Banach space $B$,
$\gamma: \mathbb{R} \rightarrow B$ has derivative $v \in B$ if,
$\forall \epsilon > 0,\, \exists c > 0$ such that
\[
\| \gamma(t) - \gamma(t_0) - (t - t_0) v \| \leq c |t - t_0|^2
\] 
for $|t - t_0| < \epsilon$.  In a Fr\'echet space, however, one has no
norm, only a metric, and if one uses the metric to try to define
derivatives, nice properties like ``sum of two differentiable vector
fields is a differentiable vector field'' will likely fail since one
does not have the nice norm properties.

Now, once one has bitten the bullet and chosen to work with the type
of space one considers the lesser of two evils, one might have thought
that all one's travail would be behind one.  Sadly, no.
\begin{theorem}
  \label{thm:no-inf-dim-spc-leb}
  The only locally finite, translation-invariant Borel measure on an
  infinite-dimensional, separable Fr\'echet space is the trivial
  measure (\emph{viz}., the one that assigns measure zero to every
  measurable set).
\end{theorem}
Thus, since a Banach space is automatically a Fr\'echet space, any
translation-invariant measure on any reasonably well behaved
infinite-dimensional space assigns infinite measure to all open sets,
unless the measure is the trivial measure.\footnote{See,
  \emph{e}.\emph{g}., \citeN{hunt-etal-trans-inv-ae-infdim-mnfld} for
  a discussion of the theory, and some interesting possible
  generalizations of relevant measure-like notions to the
  infinite-dimensional case, which I hope to discuss in future work.}

Before we abandon all hope, however, the hopeful cosmologist can now
point out that the family of Lorentz metrics on a fixed manifold is
not actually a Fr\'echet space: in particular, it is not a vector
space, since the sum of two Lorentz metrics is not in general itself a
Lorentz metric.  In fact, the family of Lorentz metrics forms a
Fr\'echet manifold, an infinite-dimensional manifold $G$ with the
local structure of a separable Fr\'echet space $F$, rather than the
local structure of $R^n$ as for an ordinary differential manifold.
(See \citeNP{geroch-inf-dim-mnflds} for a rigorous characterization of
such manifolds and a discussion of their properties.)  For such a
manifold, it is not required that the subsets that define the charts
themselves have a full vector-space structure.  (In fact, they usually
won't.)  One demands for the charts $(U_i, \phi_i)$ only that:
\begin{enumerate}
    \item the union of all $U_i$ is the entire manifold $G$
    \item each $\phi_i$ is a bijection from $U_i$ to an open subset of
  $F$ (where the topology is determined by the translation-invariant
  metric on $F$)
    \item for all $i$ and $j$, $\phi_i [U_i \cap U_j]$ is an open set
  in $F$
    \item for all $i$ and $j$, $\phi_i \circ \phi^{-1}_j [U_i \cap
  U_j]$ has continuous Fr\'echet derivatives up to whatever order one
  requires for the manifold\footnote{See \citeN{geroch-inf-dim-mnflds}
    for a definition of the Fr\'echet derivative.}
\end{enumerate}
Given any two Lorentz metrics $g_{ab}$ and $h_{ab}$ on a spacetime
manifold such that the null cones of one are contained in the other,
there is always a small enough $\epsilon$ that $g_{ab} + \epsilon
h_{ab}$ and $g_{ab} - \epsilon h_{ab}$ are also Lorentz metrics.  One
can use such sets of Lorentz metrics to construct the local Fr\'echet
structure of the manifold of all Lorentz metrics.

Now, it is a theorem that all separable Fr\'chet manifolds are
diffeomorphic to an open subset of the separable, infinite-dimensional
Hilbert space \cite{henderson-inf-dim-mnflds-hilb-spc}, itself a
Fr\'echet space.  There now follows from
theorem~\ref{thm:no-inf-dim-spc-leb}:\footnote{I do not know of a
  proof of this theorem in the literature, but it is not difficult to
  show.}
\begin{theorem}
  \label{thm:no-inf-dim-man-leb}  
  There is no locally translation-invariant Borel measure on an
  infinite-dimensional Fr\'echet manifold.
\end{theorem}
``Locally translation-invariant'' means the following: fix a point $g$
of the manifold, a chart $(O, \, \phi)$ containing the point and a
measurable neighborhood $N$ of $g$ contained in $O$; then any
translation of $N$ using the local Fr\'echet linear structure that
leaves $N$ entirely in $O$ preserves the measure of $N$.  For the
cases of interest to us, the measure should respect such ``local,
first-order translations'', because those are exactly the kinds of
perturbations that cosmologists always consider when discussing the
genericity and stability of properties of spacetimes.

\section{Topologies and Measures on Families of Spacetimes}
\label{sec:topol-meas-sts}

Even though, as should now be clear, we cannot hope for the most
satisfactory framework---a well behaved Borel measure---on which to
found probabilistic reasoning about families of spacetimes, we may
still hope to find topologies or measures on their own appropriate for
addressing specific sorts of problems.  Perhaps, the hope goes, we can
find a well behaved, physically significant measure that will return
probabilities in such a way that its lack of relation to a topology
will not necessarily lead to conundrums or implausibility.  Or perhaps
we can find a topology that, though not related to a measure, will
still allow us to reason qualitatively about likelihoods in a physically
significant way.  Alas, in the event, things do not look good.

\subsection{Topologies}
\label{sec:topo-sts}
\resetsub


Fix a candidate spacetime manifold $M$, and consider the family
$\mathfrak{G}$ of all Lorentz metrics on it, \emph{i}.\emph{e}., the
family of all cross-sections of the fiber bundle of Lorentz metrics
over $M$.  There are two standard topologies relativists impose on
$\mathfrak{G}$ when addressing problems related to likelihoods.  The
first is a standard compact-open topology on a function space; the
second is a standard Whitney topology on a function
space.\footnote{\label{fn:sobolev}\citeN{geroch-gr-in-large} calls the
  former the coarse and the latter the fine topology.
  \citeN{hawking-stab-gen-props-gr} calls the latter the open
  topology, and calls a yet finer topology the fine topology.  We
  shall not consider here the fine topology of
  \citeN{hawking-stab-gen-props-gr}, for, as we shall soon see, the
  Whitney topology already has ``too many open sets''.  Another common
  class of topologies used in relativity theory are the Sobolev
  topologies, which play an important role in the analysis of the
  Cauchy problem in general relativity
  \cite{ringstrom-cauchy-prob-gr}.  Since these are even finer than
  the finest one \citeN{hawking-stab-gen-props-gr} considers, and
  since we shall not discuss the Cauchy problem, we shall, again, not
  worry about them.}  (The compact-open is strictly coarser than the
Whitney, unless $M$ itself is compact, in which case the two coincide;
we will not consider that case.)  The idea behind each is to fix a
standard of ``distance'' between Lorentz metrics by fixing an
arbitrary positive-definite metric on $M$ and using it to assign
magnitudes to the algebraic differences of Lorentz metrics.  As we
shall see, both have severe problems of physical interpretation, which
can in large part be traced to the fact that the positive-definite
metrics themselves used to fix the similarity relations among Lorentz
metrics have no physical significance.\footnote{See
  Geroch~\citeyear{geroch-topol-gr,geroch-gr-in-large} and
  \citeN{fletcher-sim-topo-psig-rel} for insightful discussions of
  these problems.}

Roughly speaking, the compact-open topology cares only whether or not
metrics are similar on bounded regions in the interior of the
spacetime manifold; it does not care about their relative asymptotic
behavior.  To characterize it, we must define the neighborhoods of a
given Lorentz metric $g_{ab}$.  A neighborhood $\mathcal{N} (h_{ab},
\, K, \, \epsilon; \, g_{ab})$ is determined by a positive-definite
metric $h_{ab}$ on $M$, a compact subset $K$ of $M$, and a real number
$\epsilon > 0$.  A Lorentz metric $g'_{ab}$ is in the neighborhood if
and only if $h^{mn} h^{rs} (g_{mr} - g'_{mr}) (g_{ns} - g'_{ns}) <
\epsilon$ everywhere in $K$.  The family of all such neighborhoods
forms a subbase for the compact-open topology.\footnote{One can also
  form compact-open topologies that account for derivatives of the
  Lorentz metrics and how they differ, but we will not need to do so.}
The compact-open topology has the pleasant properties of being locally
compact, Hausdorff and regular (because the fiber bundle of Lorentz
metrics over $M$ is).  It also, according to
proposition~\ref{prop:c-o-topo}, is the coarsest mathematically
reasonable topology to use on $\mathfrak{G}$.

An example from \citeN{geroch-gr-in-large} shows, however, that its
physical significance is questionable at best.  Consider the sequence
of metrics on $\mathbb{R}^4$ of the form $\text{diag} (t_m, \, -1, \,
-1, \, -1)$, for $m \in \mathbb{I}^+$ (the strictly positive
integers), where $\displaystyle t_m \coloneq 1 + \frac{m}{1 + (x -
  m)^{1/2}}$, $x$ being a global Cartesian spacelike coordinate
function.  Roughly speaking, each of these metrics is essentially flat
almost everywhere except for a sharp peak of curvature around the
$t$-$y$-$z$-hypersurface defined by $x = m$.  As $m$ increases,
moreover, this peak of curvature becomes higher and sharper, as it
moves further out along the $x$-axis.  It does not seem physically
reasonable that such a sequence should converge to Minkowski
spacetime, since the spacetimes in it have curvature that in a sense
one can make precise grows without bound, and yet that is what it does
under the compact-open topology.\footnote{This sequence of metrics
  does not converge at all under the Whitney topology, which one may
  perhaps think of as the ``correct'' or ``naturally expected''
  result.}  The problem is that the compact-open topology is too
coarse---it does not have enough open sets to stop pathological
sequences from converging.

Roughly speaking, the Whitney topology cares whether or not metrics
are similar on the entire spacetime manifold, including their relative
asymptotic behavior.  A neighborhood $\mathcal{N} (h_{ab}, \,
\epsilon; \, g_{ab})$ of a given Lorentz metric $g_{ab}$ is determined
by a positive-definite metric $h_{ab}$ on $M$, and a real number
$\epsilon > 0$.  A Lorentz metric $g'_{ab}$ is in the neighborhood if
and only if $h^{mn} h^{rs} (g_{mr} - g'_{mr}) (g_{ns} - g'_{ns}) <
\epsilon$ everywhere in $M$.  The family of all such neighborhoods
forms a subbase for the topology.\footnote{One can also form Whitney
  topologies that account for derivatives of the Lorentz metrics and
  how they differ, but again we will not need to do so.}  The Whitney
topology also has the pleasant properties of being locally compact,
Hausdorff and regular (because the fiber bundle of Lorentz metrics
over $M$ is).

The Whitney topology fares even worse than the compact-open one with
regard to physical significance, as examples from
Geroch~\citeyear{geroch-sings,geroch-gr-in-large} again show.
Consider the sequence of metrics on $\mathbb{R}^4$ of the form
$\text{diag} (t_m, \, -1, \, -1, \, -1)$, for $m \in \mathbb{I}^+$,
where now $\displaystyle t_m \coloneq 1 + \frac{1}{m^2 + x^2 + y^2
  +z^2}$.  Each metric in this family is essentially flat almost
everywhere except for a spherically symmetric bump of curvature
centered on the origin; this bump, moreover, decreases smoothly to
zero as $m$ increases.  This sequence, however, does \emph{not}
converge to Minkowski spacetime under the Whitney topology.  Even more
egregiously, the one-parameter family of metrics $\{\lambda g_{ab}\}$,
for $\lambda \in \mathbb{R}^+$ (the set of strictly positive real
numbers), where $g_{ab}$ is \emph{any} Lorentz metric on \emph{any}
non-compact $M$, fails to be a continuous curve under the Whitney
metric.  But each metric in the family represents the same physical
spacetime!  Multiplying a spacetime metric by a constant does nothing
other than change the effective units one (implicitly) uses to
quantify physical magnitudes such as mass and
acceleration.\footnote{The compact-open topology seems to get both of
  these examples right: under it, the sequence converges to Minkowski
  spacetime, and the one-parameter family forms a continuous curve.}
The problem now is that the Whitney topology is too fine---it has so
many open sets that almost no reasonable sequence will converge.

One could perhaps argue with some justice that Geroch's example
speaking against the compact-open topology is not so bad as to
preclude its usefulness in many cases and for many purposes, and I
would not necessarily disagree.  The problem arises in the example
because the compact-open topology, roughly speaking, does not contain
enough open sets to control the similarity relations between metrics
with respect to their asymptotic behavior.  In other words, it does
not care about global similarity, only local similarity.  For the
sorts of problems for which one would want to use a topology on a
family of spacetime metrics to ground qualitative probabilistic
reasoning in the context of cosmology, however, it is exactly the
global similarity of metrics that will in general be at issue.  We
will see physically important examples of this in
\S\ref{sec:gener-scarc-pred} below.  The Whitney topology rules
unhelpfully in such simple and fundamental cases as to make it, to my
mind, never a viable option for addressing global issues.\footnote{One
  may thus wonder about the usefulness of the Sobolev measures used in
  analyzing the stability of the Cauchy problem in general relativity,
  for those topologies are always strictly finer even than the Whitney
  topology.  In fact, though, in this case the promiscuity of the
  topologies is a virtue---one \emph{wants} to show stability under as
  difficult conditions as possible.  Even though one may not be able
  to elucidate the physical significance of the Sobolev topologies,
  they are surely finer than any topology one will be able so to
  elucidate, and so stability under the Sobolev topologies ensures
  stability under more restrained, more physically plausible ones,
  whatever they may be.}

It will be useful to conclude the section by considering problems with
these topologies in the context of a more physically interesting
example, to substantiate my claims.  Consider the question of the
stability of the occurrence of singularities in the family of metrics
over a given manifold.  One wants to show that the occurrence of a
singularity is (topologically) stable ``under small perturbations''.
(I discuss this question, and the problems facing attempts to address
it, in more detail in \S\ref{sec:gener-scarc-pred} below.)  In this
case, the impropriety of the Whitney topology can be easily
illustrated by noting that any reasonable sense of ``small
perturbation'' will yield an operation discontinuous with respect to
it.  For example, given a spacetime $(\mathcal{M}, \, g_{ab})$, one
might define a small perturbation as follows.  Consider a
one-parameter family of spacetimes $\mathfrak{M}_\epsilon \coloneq \{
(\mathcal{M}, \, (1 + \phi_\lambda) g_{ab}) : \lambda \in [0, \,
\epsilon) \}$, for some small $\epsilon$, where each $\phi_\lambda$ is
a non-negative smooth function on $\mathcal{M}$ such that $0 \ge
\sup_\mathcal{M} \phi_{\lambda'} < \lambda' < \sup_\mathcal{M}
\phi_\lambda < \lambda$, for all $\lambda', \lambda \in [0, \,
\epsilon)$, and the family of functions $\{ \phi_\lambda \}$ varies
smoothly with respect to $\lambda$ in the supremum norm, and the
supremum approaches zero ``slowly''.\footnote{Since the family of
  Lorentz metrics is a locally convex space, all the $(1 +
  \phi_\lambda) g_{ab}$ will also be Lorentzian for small enough
  $\epsilon$.}  Then
\[
\left( 1 + \left. \frac{\text{d} \phi_\lambda} {\text{d} \lambda}
  \right|_{\lambda = 0} \right) g_{ab}
\]
is a small perturbation off $g_{ab}$.  It is easy to see by
construction that $\mathfrak{M}_\epsilon$ forms an everywhere
discontinuous curve on the family of metrics with respect to the
Whitney topology, and so \emph{any} property of $(\mathcal{M}, \,
g_{ab})$ one may want to consider is trivially stable under such small
perturbations.  (The only physically reasonable ``small perturbation''
continuous with respect to the Whitney topology is the identity
operation.)

Non-trivial small perturbations defined in this way can easily be
constructed so as to be continuous with respect to the compact-open
topology, so this looks initially more promising.  For the treatment
of singularities, though, the compact-open topology is not physically
appropriate.  If one characterizes a singularity by the existence of
incomplete, inextendible causal geodesics, then the compact-open
topology will never be able to discriminate singular from non-singular
metrics: it is only in highly pathological cases that incomplete,
inextendible geodesics are contained in compact subsets of a spacetime
\cite{curiel-sing}.  Every open neighborhood of a singular metric in
the compact-open topology contains non-singular metrics, and
vice-versa.  There are no other well defined topologies on the family
of Lorentz metrics standardly used by physicists.\footnote{In order to
  try to address such problems with the compact-open and the Whitney
  topologies, \citeN{fletcher-glob-st-sim} constructs a novel
  topology, in some ways similar to the compact-open topology but
  which yields the ``natural'' answers for Geroch's examples I
  discussed above.  It would be of great interest to determine whether
  Fletcher's topology is appropriate for the characterization of the
  stability of singularities not confined to compact subsets of
  spacetime.}

\subsection{Measures}
\label{sec:meas-sts}
\resetsub

The space of Lorentz metrics over a manifold---the family of possible
spacetime models having that as its underlying manifold---is an
infinite-dimensional Fr\'echet manifold, and so by
theorem~\ref{thm:no-inf-dim-man-leb} it has no non-trivial Borel
measure.  The only rigorously defined measure on a reasonably
interesting family of spacetimes is the Gibbons-Hawking-Stewart (GHS)
measure $\mu_\textsc{ghs}$ on ``minisuperspace'' $\Gamma$
\cite{gibbons-et-natl-meas-all-univs}.  Roughly speaking, $\Gamma$
comprises the family of initial data for FLRW spacetimes with compact
Cauchy surfaces, sourced by a minimally coupled homogeneous scalar
field.  A little more precisely, one constructs the constraint-reduced
phase space for an appropriately gauge-fixed Hamiltonian formulation
of general relativity; restricting attention to compact 3-geometries
sourced by homogeneous, minimally coupled scalar fields, one finds
that the resulting space, remarkably, simplifies to the point that it
itself is only four-dimensional.  In essence, the reduced phase space
is fully parametrized by the field-intensity $\phi$ of the scalar
field and the Hubble expansion factor $a$ on a Cauchy surface, and
their ``time-derivatives'', $\dot{\phi}$ and $\dot{a}$, off the Cauchy
surface; these quantities are constant on the Cauchy surface by
homogeneity of the spacetime.  The standardly defined Liouville
measure on this phase space (modulo a few technical difficulties that
do not concern us) is the GHS measure.  All of a sudden, things are
looking up for our eternally hopeful cosmologist who would engage in
probabilistic reasoning, at least with regard to this (admittedly
quite restricted, but still physically important) family of
spacetimes---we have a rigorously defined Borel measure on a
finite-dimensional space.  (A Liouville measure is always a Borel
measure.)  It is not long, however, before a bucket of cold water is
dashed in her face with the realization that, even though the space is
finite-dimensional and the measure is Borel, it cannot be turned into
a probability measure, for the measure it assigns the entire phase
space is infinity---$\Gamma$ is not compact.

Still, let us see whether we may not salvage something useful from
this mess.  We want to see whether the GHS measure can support any,
even weak, form of probabilistic reasoning.  Say we want to determine
whether we can meaningfully attribute a probability to the occurrence
of a physical property $X$, given the fixed reference class $\Gamma$.
Let $P_X \subset \Gamma$ be the family of spacetimes evincing $X$.
There are four cases to consider.
\begin{enumerate}
    \item $P_X$ is not measurable
    \item $\mu_\textsc{ghs} (P_X) < \infty$
    \item $\mu_\textsc{ghs} (\Gamma \setminus P_X) < \infty$
    \item $\mu_\textsc{ghs} (P_X) = \infty$ and $\mu_\textsc{ghs}
  (\Gamma \setminus P_X) = \infty$
\end{enumerate}
In the first case, we can say nothing at all, but one assumes or
stipulates or hopes or demands or pleads or dreams that physically
significant properties will not manifest such topological pathology in
their distribution across spacetimes.  In the second case, one can
unambiguously attribute a probability of zero to it, and in the third
a probability of one.  In the fourth, one can say nothing simple or
straightforward, without ambiguity, but now one does not even have the
solace of yelling at the property and demanding that it not be
pathological, as in the first case, for there is nothing pathological
about such topological behavior at all.  

There is, however, a ``natural'' schema for regularization procedures
that one can use to try to derive a finite probability in such
cases.\footnote{I follow here the exposition of
  \citeN{schiffrin-wald-meas-prob-cosmo}.}  One approximates $P_X$ by
a nested sequence of finite-measure subsets of $\Gamma$, such that the
union of the sequence is $P_X$ and the sequence of measures of the
subsets, appropriately weighted, converge to a finite value in $[0, \,
1]$:
\begin{enumerate}
    \item assume $\Gamma$ is $\sigma$-finite (\emph{i}.\emph{e}., is a
  countable union of subsets of finite measure)
    \item find ``physically appropriate'' nested sequence of subsets
  of $\Gamma$, $\{S_i\}$ ($i \in \mathbb{I}^+$), such that $\Gamma =
  \bigcup_i S_i$ and $\mu_{\textsc{ghs}} (S_i) < \infty$
    \item define Pr$(P_X)$ = $\displaystyle \lim_{i \rightarrow
    \infty} \frac{\mu_{\textsc{ghs}} (P_X \cap S_i)}
  {\mu_{\textsc{ghs}} (S_i)}$
\end{enumerate}
Minisuperspace is $\sigma$-finite, so we're off to a good start.  The
serious problem arises with the second condition: one can get pretty
much any answer one wants by judicious choice of $\{S_i\}$,
\emph{i}.\emph{e}., different regularization procedures can yield
wildly different results.  

A simple example illustrates the general form of the problem.  What is
the probability that a randomly chosen natural number is even?
\emph{Prima facie}, the question makes no sense.  Let's fix a
regularization procedure to attempt to address it.  Let $S_i$ be the
subset consisting of the first $i$ natural numbers, in their normal
ordering; then the regularization procedure yields the well defined
probability $\half$ for a natural number's being even.  Now, however,
order the natural numbers as follows, $\{1, \, 3, \, 2, \, 5, \, 7, \,
4, \ldots\}$, and again let $S_i$ be the subset consisting of the
first $i$ numbers.  This yields a well defined probability, but now it
is $\athird$.

In cosmology, the problem is nicely illustrated by attempts to
calculate the probability of inflation for spacetimes in $\Gamma$.
Using regularization procedures derived from arguments based on
(topological) stability of initial conditions yielding ``slow-roll''
inflation, \citeN{gibbons-turok-meas-prob-cosmo} deduced extremely low
probability for $N \gg 1$ $e$-foldings of inflation, whereas
\citeN{carroll-tam-uniy-evol-cosmo-tun} deduced extremely high
probability for $N \gg 1$ $e$-foldings of inflation.  Both analyses,
moreover, have strong, physically plausible justifications for the
regularization procedures they employ (\emph{i}.\emph{e}., their
choice of $\{S_i\}$).  The resolution to this seemingly paradoxical
state of affairs is that, in fixing the choice of $\{S_i\}$, they each
used a different criterion for topological stability for initial
conditions yielding inflation.  Roughly speaking,
\citeN{carroll-tam-uniy-evol-cosmo-tun} characterized topological
stability based on the behavior of spacetimes entering an inflationary
phase, whereas \citeN{gibbons-turok-meas-prob-cosmo} did so based on
spacetimes leaving the inflationary phase.  This difference naturally
leads them to consider the weight the GHS measure assigns to
physically quite different open sets in $\Gamma$.  It should therefore
be no surprise that those open sets get assigned divergent
weights.\footnote{My diagnosis of the conflict between the two
  conclusions is in some ways similar to that of
  \citeN{schiffrin-wald-meas-prob-cosmo}, but also differs in one
  important way, \emph{viz}., my emphasis on the role topological
  stability plays in their arguments in fixing the open sets whose
  measures are relevant to the problem.}

Even in the cases where one can unambiguously attribute a probability
to the occurrence of a property based on $\mu_{\textsc{ghs}}$, one
must ask about the physical significance of that probability, which,
if it indeed has any, must come from the physical significance of the
GHS measure itself, if it indeed has any.
\citeN{schiffrin-wald-meas-prob-cosmo} argue persuasively on multiple
grounds that, at best, much work must be done to justify the physical
significance of the GHS measure, and, at worst, it has none.  They
note that the standard justifications for the use of a Liouville
measure are given by arguments based on special properties of the
dynamical evolution of the system at issue and in particular on how it
equilibrates.  In particular, the arguments rely on the fact that the
amount of time the system spends in a portion of phase space is
proportional to its Liouville measure.  Those arguments, however, are
not available when:
\begin{enumerate}
    \item the system is not ergodic
    \item OR one has not waited a time much greater than the
  equilibration time after the system was prepared
    \item OR the system has a time-dependent Hamiltonian that varies
  on a timescale that is small or comparable to the equilibration time
\end{enumerate}
All of those conditions hold, however, for the canonical ``dynamics''
of that sector of general relativity represented by minisuperspace and
its Hamiltonian.  The system is not ergodic because the phase space
has infinite measure---the dynamics cannot adequately explore the
entire energy hypersurface in any finite time.  For the same reason,
there is no finite time in which the system can explore enough of the
energy hypersurface in order for one to be able to conclude that it
has satisfactorily equilibrated: the system always ``remembers'' its
initial state, which precludes true statistical equilibration.  That
the time-dependent Hamiltonian in this case varies over timescales
small compared to the equilibration time follows for the same reason.

In the face of these problems, \citeN{hollands-wald-altern-inflat} and
\citeN{schiffrin-wald-meas-prob-cosmo} conclude that the only
justification for the use of a Liouville measure in cosmology, in our
current state of knowledge, is the bare assumption of the conceit of
\citeN{penrose-sing-time-asym}, to wit, that the universe's initial
conditions were, by some appropriate process, randomly selected from a
probability distribution fixed by the Liouville measure---the
``creator'' blindly threw a dart at a dartboard whose values are
distributed according to it.
\citeN[p.~20]{schiffrin-wald-meas-prob-cosmo} drily observe that this
``has the status of an unsupported hypothesis.''  I demur.  There is
no tongue long enough and no cheek deep enough to endow this
assumption with the honorific `hypothesis'.  There is \emph{no} known
physical justification for the use of the Liouville measure in
cosmology.

\section{Genericity, Stability, and Prediction}
\label{sec:gener-scarc-pred}

\resetsec

As I already remarked above, the space of Lorentz metrics over a
manifold---the family of possible spacetime models having that as its
underlying manifold---is an infinite-dimensional Fr\'echet manifold,
and so by theorem~\ref{thm:no-inf-dim-man-leb} it has no non-trivial
Borel measure.  Standard probabilistic forms of argument in cosmology,
however, mix topological and measure-theoretic concepts and methods in
a way that depends on relations between topology and measure that are
guaranteed to obtain only for Borel measures.  In particular, those
standard forms (always implicitly) assume at least one of the
following propositions.
\begin{itemize}
    \item Fix a ``randomly selected'' spacetime with a given property;
  if ``small perturbations'' (in a topological sense) destroy that
  property, then the collection of spacetimes with that property has
  zero measure.  (The property is ``scarce''; theorems showing the
  existence of the property are ``rigid''.)
    \item Fix a ``randomly selected'' spacetime with a given property;
  if ``small perturbations'' (in a topological sense) preserve that
  property, then the collection of spacetimes with that property has
  large (or at least discernibly non-zero) measure.  (The property is
  ``generic''; theorems showing the existence of the property are not
  ``rigid''.)
    \item If the collection of spacetimes with a given property has
  large (or at least discernibly non-zero) measure (``generic''), then
  that property is topologically stable under ``small perturbations''
  (not ``rigid'').
    \item If the collection of spacetimes with a given property has
  zero measure (``scarce''), then that property is topologically
  unstable under ``small perturbations'' (``rigid'').
\end{itemize}
The probabilistic element of the conclusions can be expressed using
the idea of likelihood (in a non-technical sense).  Standard arguments
then take the following form.  Assume that the property is generic and
that observations we make indicate that the actual spacetime
approximately satisfies the conditions of an existence theorem for
that property; then the topological stability under small
perturbations entailed by genericity guarantees that the inevitable
inaccuracies and inexactitudes in the observations cannot block the
inference that the likelihood that the property obtains in the actual
universe is high; and so we conclude that the likelihood is in fact
high.  Because one does not have a Borel measure in
infinite-dimensional Fr\'echet spaces, however, none of these
propositions hold in general for the space of Lorentz metrics over a
fixed manifold.\footnote{\citeN{hawking-stab-gen-props-gr} is
  particularly clear and explicit in sketching what I just proposed as
  a typical scheme for this sort of argument, though he does not note
  the mathematical issues I focus on.}

A good example of a powerful probabilistic conclusion based on
topological reasoning dressed up in measure-theoretic clothing
pertains to the likelihood of finding singularities in a certain class
of spacetimes.  \citeN{geroch-sings-closed-uni} conjectured that
essentially all spatially closed spacetimes either have singularities
or do not satisfy the SEC, or, somewhat more precisely, that
singularities are generic and their occurence is stable in the family
of spatially closed spacetimes.\footnote{The strong energy condition
  requires that for any timelike vector $\xi^a$, $R_{mn} \xi^m \xi^n
  \geq 0$, where $R_{ab}$ is the Ricci tensor associated with the
  spacetime metric.}  One compelling way to make Geroch's conjecture
precise is given by the so-called Lorentzian splitting
theorems.\footnote{\label{fn:splitting}In order to state the most
  relevant splitting theorem, we need two definitions.  First, the
  \emph{edge} of an achronal, closed set $\Sigma$ is the set of points
  $p \in \Sigma$ such that every open neighborhood of $p$ contains a
  point $q \in I^-(p)$, a point $r \in I^+(p)$ and a timelike curve
  from $q$ to $r$ that does not intersect $\Sigma$.  Second, let
  $\Sigma$ be a non-empty subset of spacetime; then a future
  inextendible causal curve is a \emph{future $\Sigma$-ray} if it
  realizes the supremal Lorentzian distance between $\Sigma$ and any
  of its points lying to the future of $\Sigma$
  \cite{galloway-horta-reg-lorentz-busemann-fns}; \emph{mutatis
    mutandis} for a \emph{past $\Sigma$-ray}.  (If $\gamma$ is a
  $\Sigma$-ray, it necessarily intersects $\Sigma$.)
  \begin{theorem}[Lorentzian splitting theorem]
    $(\mathcal{M}, \, g_{ab})$ be a spacetime that contains a compact,
    acausal spacelike hypersurface $\Sigma$ without edge and obeys the
    SEC; if it is timelike geodesically complete and contains a future
    $\Sigma$-ray $\gamma$ and a past $\Sigma$-ray $\eta$ such that
    $I^- (\gamma) \cap I^+ (\eta) \neq \emptyset$, then it is
    isometric to $(\mathbb{R} \times \Sigma, \, t_a t_b - h_{ab})$,
    where $(\Sigma, \, h_{ab})$ is a compact Riemannian manifold and
    $t^a$ is a timelike vector-field in $\mathcal{M}$.
  \end{theorem}
  In particular, $(\mathcal{M}, \, g_{ab})$ must be globally
  hyperbolic and static.  See
  \citeN{galloway-horta-reg-lorentz-busemann-fns} for a proof.}  These
theorems may be thought of as rigidity meta-theorems for singularity
theorems invoking the strong energy condition, for the splitting
theorems show that, under certain other assumptions, there will be no
singularities only when the spacetime is static and globally
hyperbolic.\footnote{See \citeN[ch.~14]{beem-et-glob-lor-geom} for a
  beautiful discussion of the rationale behind and intent of rigidity
  theorems, as well as an exposition of many of the most important
  ones.}  The reasoning then runs, static and globally hyperbolic
spacetimes are ``of measure zero'' in the space of all spacetimes, and
so being free of singularities is, under the ancillary conditions,
unstable under arbitrarily small perturbations; thus, the likelihood
of a ``randomly selected'' spatially closed spacetime being
singularity-free is very low.\footnote{See, \emph{e}.\emph{g}.,
  \citeN{hawking-stab-gen-props-gr}, \citeN{penrose-sing-time-asym},
  and \citeN{senovilla-sing-thms-conseq} for examples of physicists
  explicitly using such measure-theoretic language to characterize the
  genericity of the occurrence of singularities in these families of
  spacetimes, based on topological stability of the occurrence of
  singularities.  Those same physicists also offer similar arguments
  for the genericity of singularities in spatially open spacetimes.
  One can make the conjecture in this case precise by using a
  variation of the Lorentzian splitting theorem given in
  footnote~\ref{fn:splitting}
  \cite{galloway-horta-reg-lorentz-busemann-fns}; see,
  \emph{e}.\emph{g}., \citeN{ringstrom-cauchy-prob-gr}, for arguments
  of the sort I criticize based on the Lorentzian theorem for the
  spatially open case.}  These conclusions, however, are simply not
justified in the absence of a Borel measure, even if one had a
physically appropriate topology to use for the rigorous
characterization of stability in the first place.


An example of a different sort is provided by the anthropic argument
of \citeN{weinberg-anthrop-bnd-cc} predicting ``the most likely''
range of values for the cosmological constant $\Lambda$.  His argument
runs as follows:
\begin{enumerate}
    \item work with a family of near-FLRW spacetimes
  (\emph{i}.\emph{e}., ones derived by allowing small perturbations
  off FLRW spacetimes, introducing small inhomogeneities);
    \item then the existence of large, gravitationally bound systems
  places upper and lower bounds on possible values of $\Lambda$---if
  $\Lambda$ is too positive, then potentially bound systems would be
  pulled apart, and if it is too negative, then the universe would
  recollapse before they can form;
    \item argue for the topological stability of the formation of such
  bound systems under small changes in the value of $\Lambda$;
    \item use an anthropic argument (the presence of conscious
  observers as a selection effect, assuming we are typical observers,
  \emph{i}.\emph{e}., that the value of $\Lambda$ in our spacetime is
  typical of spacetimes with such observers) to fix the shape and peak
  of an appropriate measure on the family of near-FLRW spacetimes;
    \item predict that the probability of the occurrence of a
  cosmological constant with a value lying in the range fixed in the
  second step is high, according to the posited measure.
\end{enumerate}
The inadmissibility of the reasoning should, again, be clear.  The
argument assumes that there exists a measure and a topology that
harmonize in such a way as to allow one both to characterize
topological stability under small perturbations and to characterize
typicality of a class of observers in a consistent way.  On any
reasonable family of near-FLRW spacetimes, however, there will be no
such measure and topology, for the inhomogeneities ensure that the
family will form an infinite-dimensional space.

My arguments do not show that the conclusions of the sorts of
arguments I have considered in this section are necessarily wrong,
only that the arguments currently given for those conclusions, in
their present form, have serious mathematical, physical and conceptual
problems that must be addressed before any real confidence can be had
in those conclusions.

\bibliographystyle{chicago}

\begin{thebibliography}{}

\bibitem[\protect\citeauthoryear{Beem, Ehrlich, and Easley}{Beem
  et~al.}{1996}]{beem-et-glob-lor-geom}
Beem, J., P.~Ehrlich, and K.~Easley (1996).
\newblock {\em Global Lorentzian Geometry\/} (Second ed.).
\newblock New York: Marcel Dekker.

\bibitem[\protect\citeauthoryear{Carroll and Tam}{Carroll and
  Tam}{2010}]{carroll-tam-uniy-evol-cosmo-tun}
Carroll, S. and H.~Tam (2010).
\newblock Unitary evolution and cosmological fine-tuning.
\newblock Available at \href{http://arxiv.org/abs/1007.1417} {arXiv:1007.1417
  [hep-th].}

\bibitem[\protect\citeauthoryear{Curiel}{Curiel}{1999}]{curiel-sing}
Curiel, E. (1999).
\newblock The analysis of singular spacetimes.
\newblock {\em Philosophy of Science\/}~{\em 66\/}(S1), 119--145.
\newblock A more recent version, with corrections and emendations, is available
  at \url{http://strangebeautiful.com/phil-phys.html}.

\bibitem[\protect\citeauthoryear{Ellis}{Ellis}{2007}]{ellis-issues-phil-cosmo}
Ellis, G. (2007).
\newblock Issues in the philosophy of cosmology.
\newblock In J.~Butterfield and J.~Earman (Eds.), {\em Handbook of Philosophy
  of Physics, Part B}, pp.\  1183--1286. Dordrecht: North Holland.
\newblock Preprint available at \href{http://arxiv.org/abs/astro-ph/0602280}
  {arXiv:astro-ph/0602280v2}.

\bibitem[\protect\citeauthoryear{Fletcher}{Fletcher}{2014}]{fletcher-glob-st-sim}
Fletcher, S. (2014).
\newblock Global spacetime similarity.
\newblock Unpublished manuscript.

\bibitem[\protect\citeauthoryear{Fletcher}{Fletcher}{2015}]{fletcher-sim-topo-psig-rel}
Fletcher, S. (2015).
\newblock Similarity, topology, and physical significance in relativity theory.
\newblock {\em British Journal for the Philosophy of Science\/}.
\newblock Forthcoming.

\bibitem[\protect\citeauthoryear{Galloway and Horta}{Galloway and
  Horta}{1996}]{galloway-horta-reg-lorentz-busemann-fns}
Galloway, G. and A.~Horta (1996, May).
\newblock Regularity of {L}orentzian {B}usemann functions.
\newblock {\em Transactions of the American Mathematical Society\/}~{\em
  348\/}(5), 2063--2084.

\bibitem[\protect\citeauthoryear{Geroch}{Geroch}{1966}]{geroch-sings-closed-uni}
Geroch, R. (1966).
\newblock Singularities in closed universes.
\newblock {\em Physical Review Letters\/}~{\em 17}, 445--447.
\newblock \href{http://dx.doi.org/10.1103/PhysRevLett.17.445}
  {doi:10.1103/PhysRevLett.17.445}.

\bibitem[\protect\citeauthoryear{Geroch}{Geroch}{1967}]{geroch-topol-gr}
Geroch, R. (1967).
\newblock Topology in general relativity.
\newblock {\em Journal of Mathematical Physics\/}~{\em 8\/}(4), 782--786.
\newblock \href{http://dx.doi.org/10.1063/1.1705276} {doi:10.1063/1.1705276}.

\bibitem[\protect\citeauthoryear{Geroch}{Geroch}{1969}]{geroch-spin-struc-gr-i}
Geroch, R. (1969).
\newblock Spinor structure of space-times in general relativity \textsc{i}.
\newblock {\em Journal of Mathematical Physics\/}~{\em 9}, 1739--1744.

\bibitem[\protect\citeauthoryear{Geroch}{Geroch}{1970}]{geroch-sings}
Geroch, R. (1970).
\newblock Singularities.
\newblock In M.~Carmeli, S.~Fickler, and L.~Witten (Eds.), {\em Relativity},
  pp.\  259--291. New York: Plenum Press.

\bibitem[\protect\citeauthoryear{Geroch}{Geroch}{1971}]{geroch-gr-in-large}
Geroch, R. (1971).
\newblock General relativity in the large.
\newblock {\em General Relativity and Gravitation\/}~{\em 2\/}(1), 61--74.

\bibitem[\protect\citeauthoryear{Geroch}{Geroch}{1975}]{geroch-inf-dim-mnflds}
Geroch, R. (1975).
\newblock Infinite-dimensional manifolds.
\newblock Unpublished manuscript (lecture notes from a course taught by
  Geroch). Available at \url{http://strangebeautiful.com/other-minds.html}.

\bibitem[\protect\citeauthoryear{Gibbons, Hawking, and Stewart}{Gibbons
  et~al.}{1987}]{gibbons-et-natl-meas-all-univs}
Gibbons, G., S.~Hawking, and J.~Stewart (1987, February).
\newblock A natural measure on the set of all universes.
\newblock {\em Nuclear Physics B\/}~{\em 281\/}(3--4), 736--751.
\newblock \href{http://dx.doi.org/10.1016/0550-3213(88)90008-9}
  {doi:10.1016/0550-3213(88)90008-9}.

\bibitem[\protect\citeauthoryear{Gibbons and Turok}{Gibbons and
  Turok}{2008}]{gibbons-turok-meas-prob-cosmo}
Gibbons, G. and N.~Turok (2008).
\newblock Measure problem in cosmology.
\newblock {\em Physical Review D\/}~{\em 77}, 063516.
\newblock \href{http://dx.doi.org/10.1103/PhysRevD.77.063516}
  {doi:10.1103/PhysRevD.77.063516}. Preprint available at
  \href{http://arxiv.org/abs/hep-th/0609095} {arXiv:hep-th/0609095}.

\bibitem[\protect\citeauthoryear{Halmos}{Halmos}{1950}]{halmos-meas-thry}
Halmos, P. (1950).
\newblock {\em Measure Theory}.
\newblock New York: Van Nostrand and Co.

\bibitem[\protect\citeauthoryear{Hawking}{Hawking}{1971}]{hawking-stab-gen-props-gr}
Hawking, S. (1971).
\newblock Stable and generic properties in general relativity.
\newblock {\em General Relativity and Gravitation\/}~{\em 1\/}(4), 393--400.

\bibitem[\protect\citeauthoryear{Henderson}{Henderson}{1969}]{henderson-inf-dim-mnflds-hilb-spc}
Henderson, D. (1969).
\newblock Infinite-dimensional manifolds are open subsets of hilbert space.
\newblock {\em Bulletin of the American Mathematical Society\/}~{\em 75},
  759--762.
\newblock \href{http://dx.doi.org/10.1090/S0002-9904-1969-12276-7}
  {doi:10.1090/S0002-9904-1969-12276-7}.

\bibitem[\protect\citeauthoryear{Hocking and Young}{Hocking and
  Young}{1988}]{hocking-young61}
Hocking, J. and G.~Young (1988).
\newblock {\em Topology}.
\newblock New York: Dover Publications, Inc.
\newblock Originally published in 1961 by Addison-Wesley Publishing Co.

\bibitem[\protect\citeauthoryear{Hollands and Wald}{Hollands and
  Wald}{2002}]{hollands-wald-altern-inflat}
Hollands, S. and R.~Wald (2002).
\newblock An alternative to inflation.
\newblock {\em General Relativity and Gravitation\/}~{\em 34}, 2043--2055.
\newblock \href{http://dx.doi.org/10.1023/A:1021175216055}
  {doi:10.1023/A:1021175216055}. Preprint available at
  \href{http://arxiv.org/abs/gr-qc/0205058} {arXiv:gr-qc/0205058}.

\bibitem[\protect\citeauthoryear{Hunt, Sauer, and Yorke}{Hunt
  et~al.}{1992}]{hunt-etal-trans-inv-ae-infdim-mnfld}
Hunt, B., T.~Sauer, and J.~Yorke (1992).
\newblock Prevalence: a translation-invariant ``almost every'' on
  infinite-dimensional spaces.
\newblock {\em Bulletin of the American Mathematical Society\/}~{\em 27},
  217--238.
\newblock \href{http://dx.doi.org/10.1090/S0273-0979-1992-00328-2}
  {doi:10.1090/S0273-0979-1992-00328-2}.

\bibitem[\protect\citeauthoryear{Kelley}{Kelley}{1955}]{kelley-gen-topo}
Kelley, J. (1955).
\newblock {\em General Topology}.
\newblock The University Series in Higher Mathematics. Princeton: D.~Van
  Nostrand Company, Inc.

\bibitem[\protect\citeauthoryear{Peirce}{Peirce}{1878a}]{peirce-doct-chnc}
Peirce, C.~S. (1878a).
\newblock The doctrine of chances.
\newblock See \citeN{peirce-sel-phil-wrt}, Chapter~10, pp.\  142--154.
\newblock Originally published in \emph{Popular Science Monthly} 12(March,
  1878):604--615.

\bibitem[\protect\citeauthoryear{Peirce}{Peirce}{1878b}]{peirce-prob-ind}
Peirce, C.~S. (1878b).
\newblock The probability of induction.
\newblock See \citeN{peirce-sel-phil-wrt}, Chapter~10, pp.\  155--69.
\newblock Originally published in \emph{Popular Science Monthly} 12(April,
  1878):705--18.

\bibitem[\protect\citeauthoryear{Peirce}{Peirce}{1992}]{peirce-sel-phil-wrt}
Peirce, C.~S. (1992).
\newblock {\em The Essential Peirce: Selected Philosophical Writings}, Volume 1
  (1867--1893).
\newblock Bloomington, IN: Indiana University Press.

\bibitem[\protect\citeauthoryear{Penrose}{Penrose}{1979}]{penrose-sing-time-asym}
Penrose, R. (1979).
\newblock Singularities and time-asymmetry.
\newblock In S.~Hawking and W.~Israel (Eds.), {\em General Relativity: An
  {E}instein Centenary Survey}, pp.\  581--638. Cambridge University Press.

\bibitem[\protect\citeauthoryear{Ringstr\"om}{Ringstr\"om}{2009}]{ringstrom-cauchy-prob-gr}
Ringstr\"om, H. (2009).
\newblock {\em The Cauchy Problem in General Relativity}.
\newblock ESI Lectures in Mathematics and Physics. Z\"urich: European
  Mathematical Society Publishing House.

\bibitem[\protect\citeauthoryear{Schiffrin and Wald}{Schiffrin and
  Wald}{2012}]{schiffrin-wald-meas-prob-cosmo}
Schiffrin, J. and R.~Wald (2012).
\newblock Measure and probability in cosmology.
\newblock \href{http://xxx.lanl.gov/abs/1202.1818} {arXiv:1202.1818 [gr-qc]}.

\bibitem[\protect\citeauthoryear{Senovilla}{Senovilla}{1998}]{senovilla-sing-thms-conseq}
Senovilla, J. (1998).
\newblock Singularity theorems and their consequences.
\newblock {\em General Relativity and Gravitation\/}~{\em 30\/}(5), 701--848.
\newblock \href{http://dx.doi.org/10.1023/A:1018801101244}
  {doi:10.1023/A:1018801101244}. \emph{N.b}.: in the article itself (as,
  \emph{e}.\emph{g}., downloaded from the \emph{General Relativity and
  Gravitation} website), the paper is indicated as appearing in volume 29,
  1997; as the author of the paper has assured me, this is erroneous.

\bibitem[\protect\citeauthoryear{Smeenk}{Smeenk}{2012a}]{smeenk-logic-cosmo-revis}
Smeenk, C. (2012a).
\newblock The logic of cosmology revisited.
\newblock Unpublished manuscript. Available on request from the author.

\bibitem[\protect\citeauthoryear{Smeenk}{Smeenk}{2012b}]{smeenk-phil-cosmo}
Smeenk, C. (2012b).
\newblock Philosophy of cosmology.
\newblock Unpublished manuscript. Available on request from the author.

\bibitem[\protect\citeauthoryear{Weinberg}{Weinberg}{1987}]{weinberg-anthrop-bnd-cc}
Weinberg, S. (1987).
\newblock Anthropic bound on the cosmological constant.
\newblock {\em Physical Review Letters\/}~{\em 59\/}(November), 2607.
\newblock \href{http://dx.doi.org/10.1103/PhysRevLett.59.2607}
  {doi:10.1103/PhysRevLett.59.2607}.

\end{thebibliography}

\end{document}